\documentclass[11pt,twoside]{article}

\usepackage{asp2006}
\usepackage{epsf}
\usepackage{psfig}
\usepackage{lscape}

\newcommand{\etal}{{et al.}}
\newcommand{\eg}{{\it e.g.,}}

\newcommand{\kms}{{km s$^{-1}$}}

\begin{document}
\title{Integral Field Spectroscopy of Extended Emission-Line Regions
around QSOs}   %%% Fill in title
%\title{QSO Extended Emission-Line Regions --- A 3D View}
\author{Hai Fu and Alan Stockton}   %%% Fill in author names
\affil{Institute for Astronomy, University of Hawaii, 2680 Woodlawn Drive, Honolulu, HI 96822, USA}    %%% Fill in author affiliations

\begin{abstract} %%% Abstract to run on from here.
Luminous extended emission-line regions (EELRs) on kpc scales surround a substantial fraction of steep-spectrum radio-loud QSOs. Although their existence has been known for over three decades, there are still major uncertainties on the physical processes responsible for their complex morphology and kinematics. We are obtaining deep integral field spectroscopy for a sample of EELRs around QSOs at z$\leq$0.5 with the Integral Field Unit (IFU) of the GMOS on the Gemini North telescope, aiming at extracting accurate kinematics of the EELRs, measuring important physical parameters (\eg\ density, temperature, metallicity) and reliable intensity ratios of diagnostic emission lines from individual clouds that comprise an EELR.  
Here we present results from the observations of the EELR of quasar 4C\,37.43. 
We show maps of gas kinematics measured from the [O\,{\sc iii}]\ $\lambda$5007
line and line-ratio diagnostic diagrams comparing the data with predictions from ionization models. We find that the ionized gas shows rather complex global kinematics, while linear velocity gradients are often seen in individual clouds. Pure photoionization by the QSO continuum is the most likely ionization mechanism for most of the EELR clouds.

\end{abstract}

%%% MAIN BODY OF TEXT GOES HERE. CONSULT "INSTRUCTIONS FOR AUTHORS USING
%%% LATEX2E MARKUP", SECTIONS 2.3-2.6 FOR HELP WITH EQUATIONS, FIGURES,
%%% AND TABLES.

%\section{Introduction}   %%% Top level section head (remove "%" symbol)
%\subsection{}   %%% Second level section head (remove "%" symbol)
%\subsubsection{}   %%% Lowest level section head (remove "%" symbol)
%\section*{}    %%% Unnumbered top level section head (remove "%" symbol)
%\subsection*{}   %%% Unnumbered second level section head (remove "%" symbol)
About one third of low-redshift steep-spectrum radio-loud quasars 
are surrounded by massive ionized nebulae that often extend to several tens of kpc from the nucleus. These extended emission-line regions (EELRs) are important because they may reflect a key stage in a quasar's life, when it violently expels gas from its host galaxy \citep{sto02,sto06,fu06}. Integral field spectrographs on large telescopes are powerful tools to study the EELRs. Here we present results from our recent study of the 4C\,37.43 EELR, which is the most luminous EELR known among QSOs at z$\leq$0.5.  
%Although at this stage it is still early to draw a conclusion, accumulating evidence suggests that there is a connection between these extended emission-line regions (EELRs) and the feedback of quasars \citep{sto02,sto06,fu06}. If the EELRs are indeed a direct result of quasar superwinds, they offer invaluable laboratories for the study of this key stage in a quasar's life. At the same time, the physical processes that control these massive clouds itself is an important question yet to be addressed. 

%We are observing a sample of QSO EELRs at z$\leq$0.5 (drawn from \citealt{sto87}) with the GMOS Integral Field Unit (IFU) on the Gemini North telescope, complemented by long slit observations. Recently we reported a detailed IFS investigation of the EELR of 3C\,249.1 (z = 0.31; \citealt{fu06}). 

\begin{figure*}[!t]
\plotone{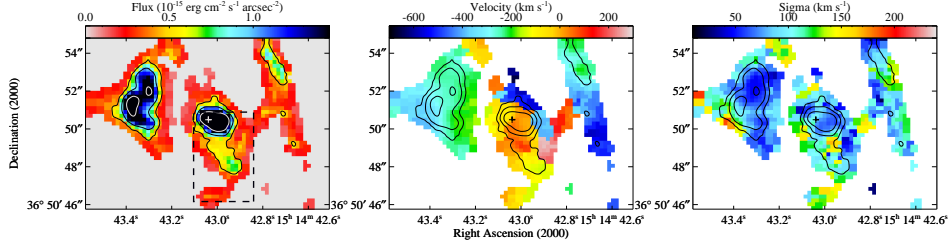}
%\plotone{colorfigs/Hai_Fu_fig1.eps}
\caption{Velocity field of 4C\,37.43 EELR derived from the
[O\,{\sc iii}]\ $\lambda$5007 line. 
%From left to right are line intensity, radial velocity
%(relative to that of the nuclear narrow line region) and velocity dispersion
%maps. 
When multiple velocities are present along the same line-of-sight, 
only the component with the lowest velocity is shown. Pixels are 0\farcs2 squares. 
The crosses indicate the position of the quasar, which has been
removed from the datacube using a PLUCY-based PSF subtraction technique. The dashed rectangle delineates the FOV of the IFUR datacube.} 
\label{fig:vfield}
\end{figure*}

4C\,37.43 was observed with the GMOS/IFU in two configurations in the early half-nights of 2006 May 23 and 24 (UT): (1) Twelve exposures of 720 s were obtained with R831 %/G5302
grating and the full-field mode (IFU2) to cover H$\beta$ and [O\,{\sc iii}]\ $\lambda\lambda$4959,5007 lines.
%He\,{\sc ii}\ $\lambda$5686 to [O\,{\sc iii}]\ $\lambda$5007; 
The telescope was dithered on a rectangular grid of $6\arcsec \times 4\arcsec$ centered on the QSO, so the final FOV is $13\arcsec \times 9\arcsec$. 
%The final mosaicked datacube (x,y,$\lambda$) has a FOV of $12\farcs8 \times 9\arcsec$ and a FWHM of $\sim0\farcs4$. 
(2) Five 2400-s exposures were taken with B600 %/G5303
grating using the half-field mode (IFUR) to win a wider spectral coverage by losing half of the field (FOV shown in Fig.~\ref{fig:vfield}). The resulting datacube includes emission lines from [Ne\,{\sc v}]\ $\lambda$3425 to [O\,{\sc iii}]\ $\lambda$5007. %The smaller FOV ($3\farcs2 \times 4\farcs7$) is centered south of the QSO (see Fig.~\ref{fig:vfield}).  
%The final datacube has a FOV of $3\farcs2 \times 4\farcs7$ and a FWHM resolution of $\sim0\farcs6$ (see Fig.~\ref{fig:vfield}). 
%We also obtained additional Keck II/DEIMOS long slit spectroscopy on two EELR clouds that are presumably associated with X-ray emissions (slit position shown in Fig.~\ref{fig:vfield}). The total integration time is 3600 s. With a 600 groove mm$^{-1}$ grating, a $0\farcs9$ wide slit and a central wavelength at 7020\AA\, it covers a spectra region from [Ne\,{\sc v}]\ $\lambda$3346 to [S\,{\sc ii}]\ $\lambda$6733. The seeing was $\sim0\farcs7$.

\begin{figure*}[!t]
  \begin{center}
    \begin{tabular}{ccc}
      \resizebox{40mm}{!}{\includegraphics{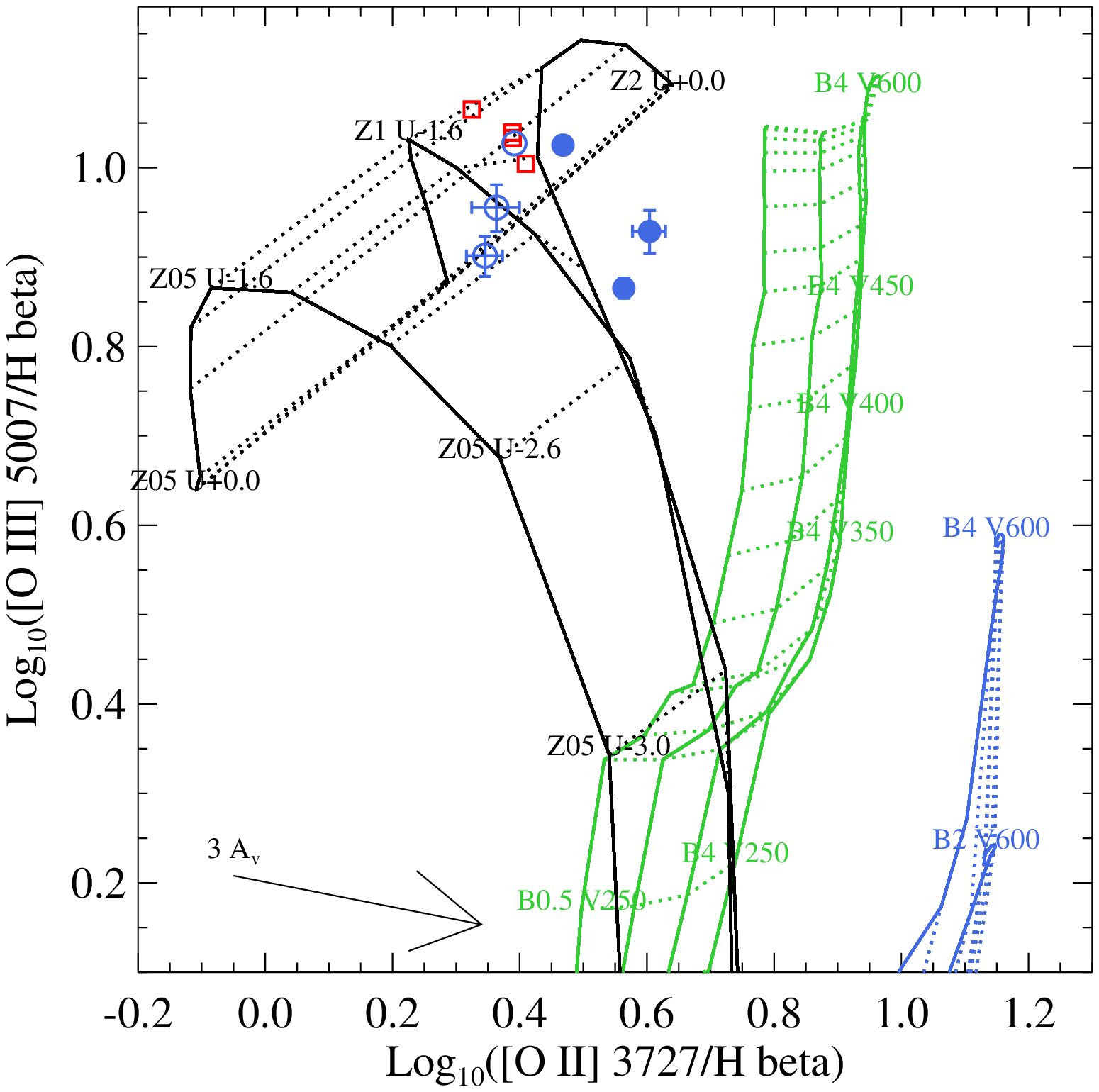}} &
      \resizebox{40mm}{!}{\includegraphics{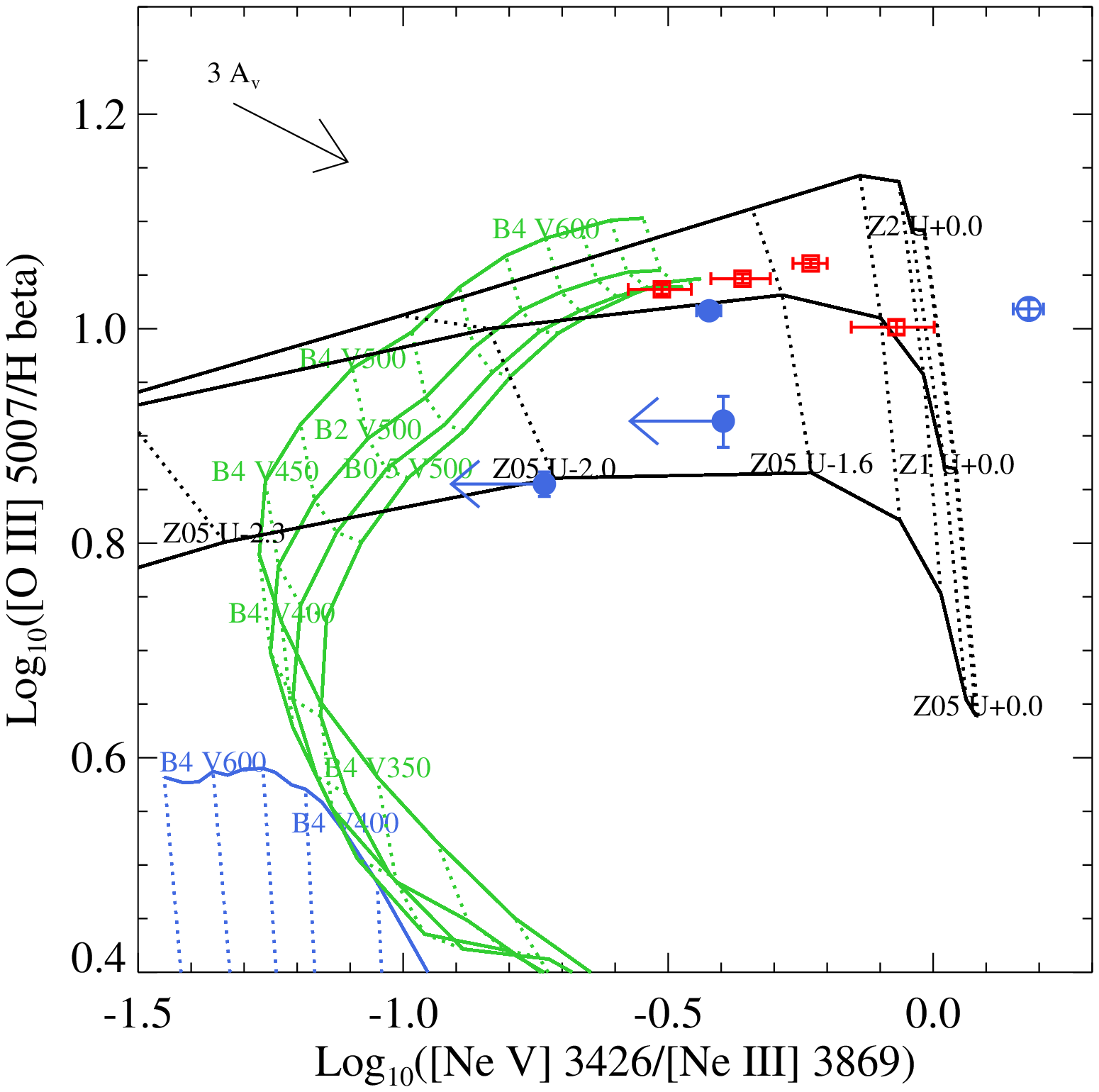}} &
      \resizebox{40mm}{!}{\includegraphics{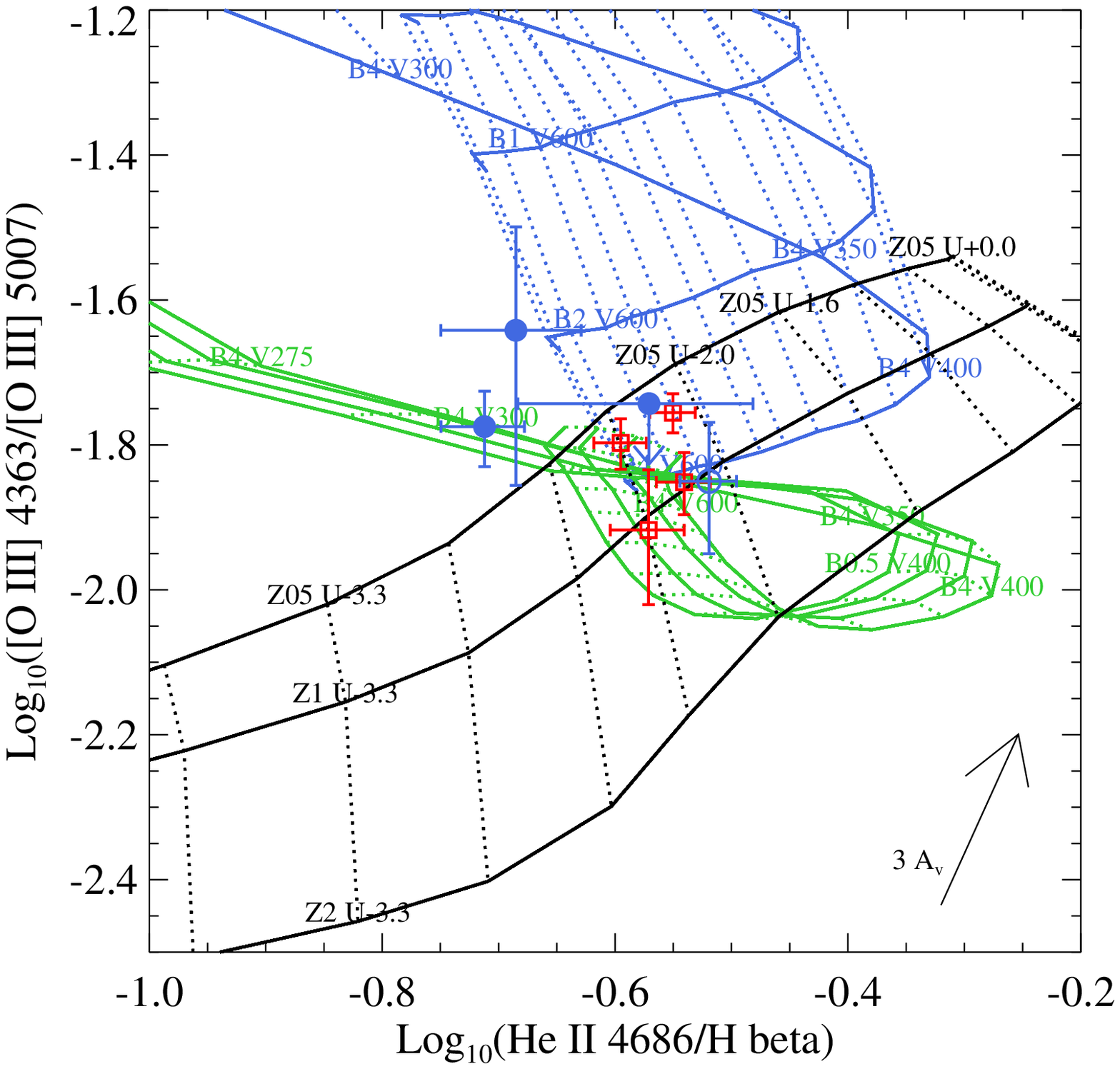}} \\
    \end{tabular}
\caption{Line-ratio diagrams are overplotted with model grids. The
pure photoionization model (Groves \etal\
2004), shock-only and ``shock + precursor" models (Dopita \& Sutherland
1996) are plotted in black, blue and green, respectively. 
Measurements from various regions around 4C\,37.43 are shown as blue circles, and red squares are data from four EELR clouds of 3C\,249.1.
Arrows are reddening vectors of 3 A$_V$.} \label{fig:diag}
 \end{center}
\end{figure*}

The IFU datacubes (x,y,$\lambda$) carry a large amount of information. Due to the space limit, we discuss only the gas kinematics and the line-ratio diagnostics here. 
The full analysis will be presented elsewhere (Fu \& Stockton, in prep.). 
(1) Gas kinematics were measured from the relatively isolated [O\,{\sc iii}]\ $\lambda$5007 line in the mosaicked IFU2 datacube. As shown in Fig.~\ref{fig:vfield}, the EELR exhibits rather complex global kinematics. The relative line-of-sight velocities range from $-$700 to +300 \kms, and obviously most of the luminous clouds are blueshifted. The velocity dispersions of most clouds are between 50 and 120 \kms.  
(2) The major ionization mechanism at work can be constrained by diagnostic diagrams involving ratios of some optical lines. Figure~\ref{fig:diag} shows examples of such diagrams. Line fluxes were measured from dereddened spectra (both Galactic and intrinsic reddenings have been taken into account), which were extracted either from the QSO-removed IFUR datacube or long slit spectra from Keck LRIS or DEIMOS. The associated 1-$\sigma$ errors were obtained using a Monte-Carlo approach. These diagrams clearly show that the pure photoionization model best fits the data.     
%(3) Luminosity-weighted average electron density can be derived from the ratio of the [O\,{\sc ii}]\ $\lambda\lambda$3726, 3729 doublet or the [S\,{\sc ii}]\ $\lambda\lambda$6716, 6731 doublet. And, electron temperature can be determined from the [O\,{\sc iii}]\
%$\lambda$4363/($\lambda$4959+$\lambda$5007) intensity ratio (\eg\
%Osterbrock 1989). Once the density is determined and the H$\beta$ flux measured, one can derive the total mass of individual clouds, and compare it with the virial mass implied by the line width (velocity dispersion).

%\acknowledgements %%% Text of acknowledgements runs on after this command.
%Based on observations obtained at the Gemini
%Observatory, which is operated by the Association of Universities for
%Research in Astronomy, Inc., under a cooperative agreement with the NSF
%on behalf of the Gemini partnership: the National Science Foundation
%(United States), the Particle Physics and Astronomy Research Council
%(United Kingdom), the National Research Council (Canada), CONICYT
%(Chile), the Australian Research Council (Australia), CNPq (Brazil) and
%CONICET (Argentina).
%This research has been partially supported by NSF grant AST03-07335.

%%% THE BIBLIOGRAPHY
%%%
%%% CONSULT SECTION 3 OF "INSTRUCTIONS FOR AUTHORS" FOR HOW TO USE NATBIB.
%%% AUTHORS ARE ENCOURAGED TO USE EITHER THE "THEBIBLIOGRAPY" ENVIRONMENT
%%% BY UNCOMMENTING (DELETING THE "%" SYMBOL) THE COMMANDS BELOW, OR BY
%%% USING THE BIBTEX ENVIRONMENT. TO FIND OUT WHICH IS APPLICABLE TO YOUR
%%% CONTRIBUTION, CONSULT THE VOLUME EDITORS FOR YOUR PROCEEDINGS.
%%%

\end{document}